\documentclass[fleqn,usenatbib]{rasti}

\usepackage{newtxtext,newtxmath}

\usepackage[T1]{fontenc}

\DeclareRobustCommand{\VAN}[3]{#2}
\let\VANthebibliography\thebibliography
\def\thebibliography{\DeclareRobustCommand{\VAN}[3]{##3}\VANthebibliography}

\usepackage{graphicx}	

\usepackage{amsmath}	
\usepackage{amssymb}	
\usepackage{siunitx}	
\usepackage{cleveref}   
\usepackage{physics}	
\usepackage[all]{nowidow}

\title[Photon counting II in the blue at a 0.5\,m telescope]{Photon counting intensity interferometry in the blue at a 0.5\,m telescope}

\author[S. Karl et al.]{
Sebastian Karl,$^{1}$\thanks{E-mail: seb.karl@fau.de}
Stefan Richter,$^{1,2}$
Joachim von Zanthier$^{1,2}$
\\
$^{1}$Friedrich-Alexander-Universität Erlangen-Nürnberg (FAU), Quantum Optics and Quantum Information, Staudtstr. 1, 91058 Erlangen, Germany\\
$^{2}$Friedrich-Alexander-Universität Erlangen-Nürnberg (FAU), Erlangen  Graduate  School  in  Advanced  Optical  Technologies  (SAOT), \\ Paul-Gordan-Str. 6, 91052 Erlangen, Germany\\
}

\date{Accepted XXX. Received YYY; in original form ZZZ}

\pubyear{2023}

\begin{document}
\label{firstpage}
\pagerange{\pageref{firstpage}--\pageref{lastpage}}
\maketitle

\begin{abstract}
Intensity interferometry is a re-emerging interferometry tool that alleviates some of the challenges of amplitude interferometry at the cost of reduced sensitivity. We demonstrate the feasibility of intensity interferometry with fast single photon counting detectors at small telescopes by utilising a telescope of diameter of merely \SI{0.5}{\m}. The entire measurement setup, including collimation, optical filtering, and two single photon detectors, is attached directly to the telescope without the use of optical fibres, facilitated by the large area of our single photon detectors. For digitisation and timing, we utilise a Time-To-Amplitude-Converter. Observing $\alpha$ Lyrae (Vega) for a total exposure time of \SI{32.4}{\hour} over the course of six nights, an auto-correlation signal with a contrast of \num[separate-uncertainty = true]{9.5(27)e-3} and a coherence time of  \SI[separate-uncertainty = true]{0.34 \pm 0.12}{\pico \s} at a SNR of 2.8 is measured. The result fits well to preceding laboratory tests as well as expectations calculated from the optical and electronic characteristics of our measurement setup. This measurement, to our knowledge, constitutes the first time that a bunching signal with starlight was measured in the B band with single photon counting detectors. Simultaneously, this is to date the stellar intensity interferometry measurement utilising the smallest telescope. Our successful measurement shows that intensity interferometry can be adopted not only at large scale facilities, but also at readily available and inexpensive smaller telescopes.
\end{abstract}




\section{Introduction}

For image formation, direct astronomical imaging as well as amplitude interferometry utilising arrays of telescopes need to preserve phase coherence throughout the optical setup \citep{Monnier2003}. This condition can be kept in both cases even at turbulent atmospheric conditions employing adaptive optics and appropriate delay lines, yet at the cost of considerable technological complexity and strain. Nonetheless, amplitude interferometry has been fruitfully implemented at interferometers like VLTI and CHARA with baselines up to \SI{127}{\m} \citep{Haubois2021} and \SI{331}{\m} \citep{Brummelaar2005}, respectively.

Intensity interferometry, on the other hand,  correlating light intensities rather than amplitudes, offers a way to avoid some of the technical challenges of amplitude interferometry. Intensity interferometry's relative simplicity in principle facilitates baselines much larger than the ones of current amplitude interferometers, up to a few kilometres  (see e.g., \cite{Dravins2013}). 
Intensity correlation measurements of starlight were first carried out with telescopes by \cite{HanburyBrown1956}, determining the diameter of Sirius to a then unprecedented precision. Expanding on previous work, \cite{HanburyBrown1974a} measured the diameter of in total 32 stars utilising the Narrabri Stellar Intensity Interferometer. Later, intensity interferometry was abandoned in favour of the more sensitive amplitude interferometers, first demonstrated at visible wavelengths by \cite{Labeyrie1975}.

In recent years, interest in intensity interferometry for astronomy has resurged due to advances in photon detection technology. For the first time since the decommissioning of the Narrabri Stellar Intensity Interferometer, temporal \citep{Guerin2017} and spatial \citep{Guerin2018}) intensity correlations of light from three bright stars were measured at the C2PU facility on the Plateau de Calern site of Observatoire de la Côte d’Azur. At the same observatory the distance to P Cygni and Rigol was estimated by combining intensity interferometry and spectroscopy \citep{Rivet2020, delAlmeida2022}. Further temporal and spatial intensity correlations were carried out since 2017, at telescopes and with arrays varying widely in size and geometry. utilising two telescopes of diameter \SI{1.8}{\m} and \SI{1.2}{\m}, respectively, and mounting the AQUEYE+ and IQUEYE instruments, \cite{Zampieri2021} measured spatial and temporal intensity correlations in the light of $\alpha$ Lyrae (Vega), at zero baseline and with a fixed baseline of over \SI{1}{\kilo \m}. Correlating light from up to three \SI{0.6}{\m} Dobsonian telescopes, \cite{Horch2021} performed spatial intensity correlation measurements for five bright stars. While all experiments mentioned so far employ conventional optical imaging telescopes, \cite{Acciari2019} measured intensity correlations for three bright stars using the two telescopes of the MAGIC array, belonging to the class of Imaging Atmospheric Cherenkov Telescops (IACTs). Performing intensity interferometry with an array of IACTs has been intensively studied in the past and is seen as a promising route for constructing very sensitive modern intensity interferometers \citep{Dravins2013}. In a first step towards this goal, all four telescopes of the VERITAS array were used to accurately measure stellar diameters with intensity interferometry for the first time since the Narrabri Intensity Interferometer by \cite{Abeysekara2020}. 

Other than the results described by \cite{Horch2021}, all intensity interferometry experiments referenced so far have been performed at larger-scale astronomical facilities. Modern single photon counting equipment however allows for intensity interferometry at smaller, commercially available telescopes, reducing the cost of an intensity interferometer greatly. We report on successful temporal intensity correlation measurements of Vega in the B band taken with a commercially available \SI{0.5}{\m} diameter telescope. 
In this paper we summarise briefly the theory of intensity correlations in \cref{sec:g2}. We then present our experimental setup including the optical design and the photon detection unit in \cref{sec:exp_setup}. Next, we briefly discuss a laboratory test of our setup in \cref{sec:lab_test}, before we finally present the results of our intensity correlation measurements obtained with a \SI{0.5}{\m} telescope observing Vega in \cref{sec:meas}. 

\section{Second order intensity correlations}
\label{sec:g2}
The second order correlation function 
\begin{equation}
    g^{(2)}(\tau) = \frac{\expval{I(t) I(t + \tau)}}{\expval{I(t)}^2} = 1 + \beta \, \abs{g^{(1)}(\tau)}^2
    \label{eq:g2}
\end{equation}
is the observable of interest in temporal intensity interferometry \citep{Mandel1995}. Here, $I(t)$ corresponds to the time dependent instantaneous intensity, the brackets denote a time average, $\beta = \frac{1}{N}$ accounts for the loss of coherence due to averaging over $N$ spatial, spectroscopic, or polarization modes, and $g^{(1)}(\tau)$ denotes the first order correlation function, sometimes called the complex degree of coherence. The second part of \cref{eq:g2}, relating first and second order correlation functions and thus field and intensity correlations, is called the Siegert relation and holds only for thermal light sources (TLS), e.g. for stars \citep{Siegert1943}. The first order correlation function is related to the normalized spectrum $s(\omega)$ of the source via the Wiener-Khintchine theorem \citep{Mandel1995}: 
\begin{equation}
    g^{(1)}(\tau) = \int_{-\infty}^\infty s(\omega) \mathrm{e}^{-i \omega \tau} \dd{\omega}
    \label{eq:wct}
\end{equation}

For a TLS, the second order correlation function displays a peak with $g^{(2)}(\tau=0) > 1$, centered around zero time delay, dropping to a baseline value of $1$ for a time delay $\tau \rightarrow \infty$. The width of the correlation peak is proportional to the source coherence time $\tau_c$. Within the coherence time, a TLS thus displays bunching, meaning it is more likely to detect two photons arriving at the detector coincidentally rather than separated by times $\tau > 0$. While in the field of quantum optics, one would conventionally define the coherence time $\tau_c$ as the inverse spectral bandwidth, we choose a definition related to the correlation peak integral:
\begin{equation}
    \tau_c = \int_{-\infty}^\infty \abs{g^{(1)}(\tau)}^2  \dd{\tau} \approx \frac{\lambda^2}{c \, \Delta \lambda}. 
    \label{eq:coh_time_def}
\end{equation}
In case of a box-transmission filter \cref{eq:coh_time_def} holds rigorously. By comparing the coherence time with the timing resolution $\Delta t$ of the detection setup and accounting for the 2 unfiltered polarization modes we can roughly estimate the peak height of the second order correlation measurement to $\nu = \frac{\tau_c}{2  \, \Delta t}$.

The signal to noise ratio of an intensity correlation measurement and its dependency of the different parameters involved has been derived by \cite{Brown1957}. These results hold even for single photon counting detectors if the photon rate is approximately constant. However, since in our experiment the photon rates at the telescope fluctuated strongly, we will not make use of this approach. Instead we calculate the theoretical noise limit to be the shot noise limit of a Poisson distributed random variable. That is, for $M$ coincidences per histogram bin, we expect a root mean square error (RMSE) of $\sqrt{M}$.

\section{Experimental Setup}
\label{sec:exp_setup}
\subsection{Telescope and optical setup}
For our experiment, we used a commercially available \SI{0.5}{\m} diameter Corrected Dall-Kirkham telescope (Planewave CDK 20). The telescope is located at the Dr. Karl Remeis-Observatory of the Astronomical Institute of the University of Erlangen-Nürnberg, located in the city of Bamberg at latitude 49°53'04.4" North and longitude 10°53'17.3" East. The telescope has a primary mirror diameter of \SI{508}{\mm}, a secondary mirror obstruction of $15.2 \%$ of the primary mirror aperture, and a focal ratio of $f/6.8$. We mounted an integrated rotating focuser directly to the telescope providing a travel range of \SI{30}{\mm}. This rotating focuser allows for the direct attachment of a \SI{9}{\kilo \g} load. The telescope mirror efficiencies are specified by the manufacturer to 96 percent, whereas no data for the efficiency of the integrated focuser could be obtained. Guiding is performed by a secondary refractive telescope attached to the telescope tubus. 

\begin{figure}
    \centering
    \includegraphics[width=0.9\columnwidth]{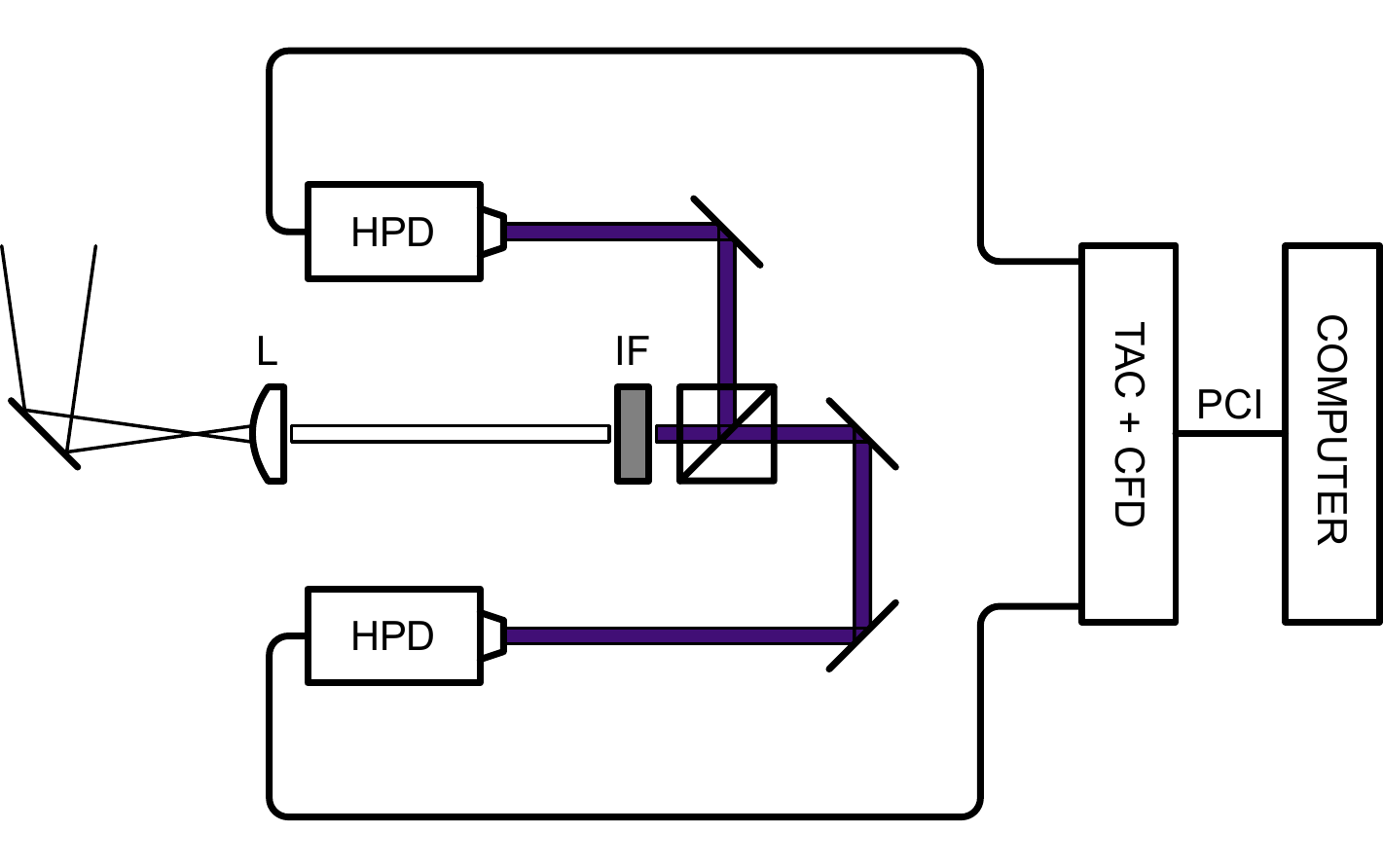}
    \caption{Sketch of the optical setup attached to the directly to the telesecope's focuser, showing all optical parts. The collimation lens is denoted as L, the interference filter is denoted as IF, and the hybrid detectors are denoted as HPD. The generated electrical pulses are discriminated by a constant-fraction-discriminator (CFD) and digitized by a time-to-amplitude-converter (TAC). The recorded histograms are transferred to a computer via PCI.}
    \label{fig:setup_sketch}
\end{figure}

The optical setup attached to the telescope is sketched in \cref{fig:setup_sketch}. After deflection into the setup plane by a protected aluminium mirror with a reflectivity of 85.5 percent (Thorlabs PF10-03-G01; all mirrors in the setup are of this type), the light is collimated by an achromatic lens (Thorlabs AC127-019-A), and filtered by an ultra-narrow interference filter with central wavelength $\lambda_0 = \SI{416.06}{\nano \m}$ and bandwidth $\Delta \lambda = \SI{0.5}{\nano \m}$ manufactured by Alluxa. The interference filter has a OD6 blocking level covering the entire visible spectrum and a peak transmission of 92 percent. After the interference filter, the light is split in two beams by a 50:50 nonpolarizing cube beam splitter (Thorlabs CCM1-BS013/M, transmissivity $> 94 \%$) in order to allow for correlation measurements with detectors whose dead time is orders of magnitude larger than the coherence time of the spectrally filtered light. The two light beams are then guided to the single photon detectors by one and two mirrors, respectively. We use single photon counting hybrid photo detectors (HPDs) of the type Becker\&Hickl HPM 100-06 with an active area of \SI{6}{\mm}, which are described in greater detail in the following subsection. After the spectral filter, all light paths are shielded by custom lens tubes to avoid stray light. This way, we can suppress the stray light to a level of less than \SI{250}{\hertz} when pointing the telescope slightly off the currently tracked star. We thus facilitate excellent background suppression without the use of an optical pinhole, considering the bright night sky directly above the city of Bamberg.

\begin{figure}
    \centering
    \includegraphics[width=0.9\columnwidth]{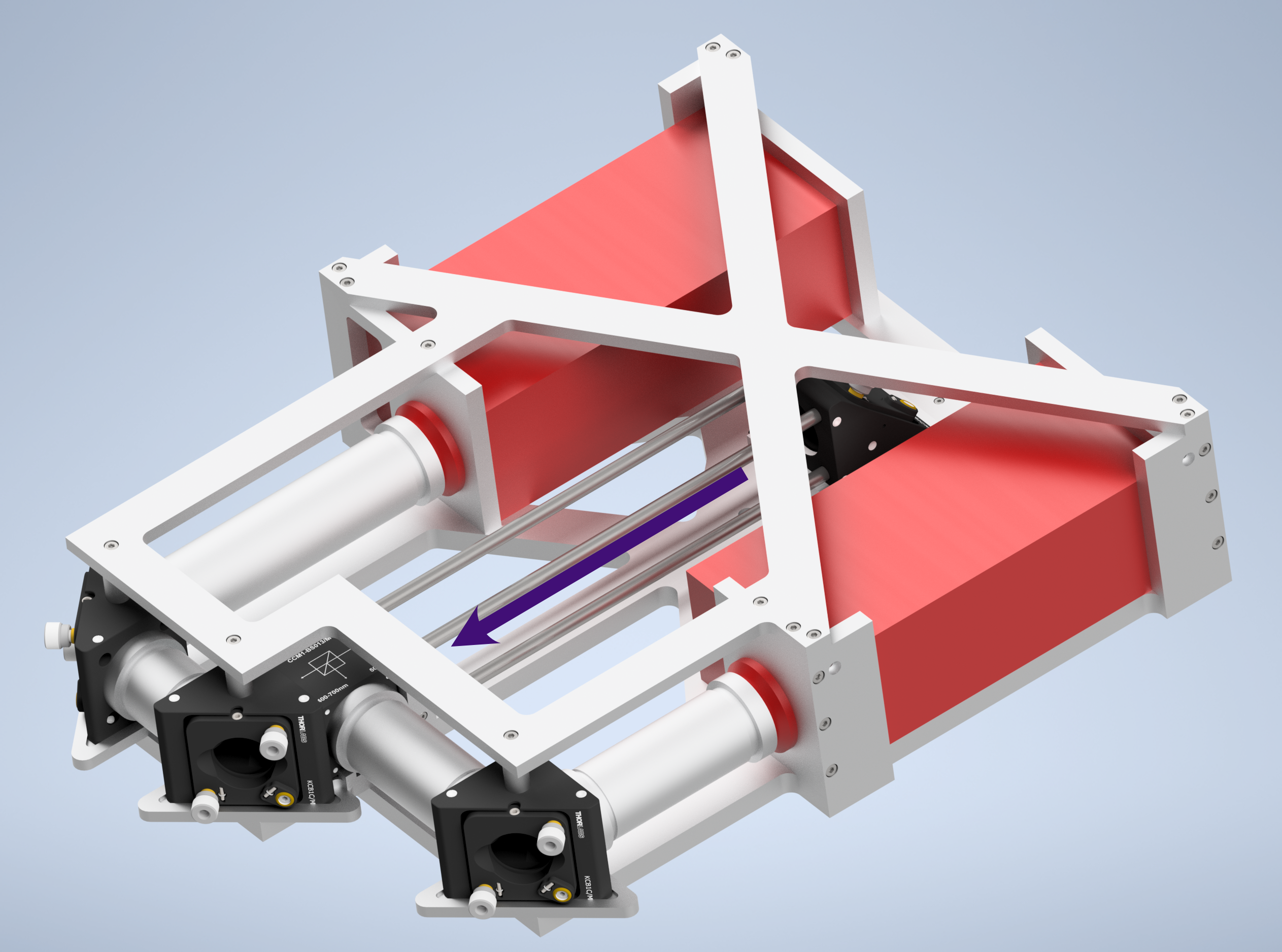}
    \caption{CAD render of the telescope attachment. The HPDs are highlighted in red, and the mirror mounts and beam splitter are shown in black. Light from the telescope would enter the setup via the rightmost mirror mount, being reflected into the cage system and hitting the beam splitter shown as a black cube to the left. From the beam splitter the light travels towards the HPDs via mirrors and custom lens tubes. The violet arrow illustrated the light's travel direction from the first deflecting mirror to the beam splitter.}
    \label{fig:setup_render}
\end{figure}

A CAD render of the telescope attachment is shown in \cref{fig:setup_render}. In this render, the light from the telescope enters the setup on the right and from below, and is reflected towards the beam splitter by the rightmost mirror mount shown in black. Before hitting the beam splitter, the light passes through a standard \SI{30}{\milli \m} cage system, in which the collimating lens and interference filter are placed. We indicate the travel direction of the light in the cage system by a violet arrow. The light then passes to the HPDs via the other mirror mounts and custom lens tubes which can be taken out and inserted while leaving all other components in place. Including all components, the telescope attachment weighs about \SI{7.7}{\kg}, keeping well within the telescope and focuser load limits.

\subsection{Photon detection equipment}
We use HPDs of the type Becker\&Hickl HPM 100-06 which combine a large active area with high timing resolution at reasonable quantum efficiencies in the blue. We employ a Becker\&Hickl SPC-130-EMN digitisation card to implement the photon arrival time correlation. Since the HPDs electric pulse height varies considerably from pulse to pulse, regular constant threshold discretising time to digital converters (TDCs) cannot be used for timing signal acquisition. Instead, correlation histograms are recorded using constant-fraction discriminators (CFDs) for signal conditioning and a time to amplitude converter (TAC). A TAC setup gives immediate access to the arrival time difference between single photons and sorts them into corresponding bins to form a correlation histogram. For our detection electronics, the channel starting the TAC is referred to as the CFD channel, while the channel stopping the TAC is referred to as the SYNC channel. Employing an SPC-130-EMN digitisation card, there is no immediate access to the SYNC or CFD channel count rates. Thus, we utilise the TAC analog-to-digital converter (ADC) readout rate for this purpose, which is equivalent to the coincidence rate over the full TAC coincidence window $\Delta \tau = \SI{40}{ \nano \s}$. To approximate the photon rates incident on the detector, we make use of the fact that the coincidence rate $n_{\mathrm{coinc}}$ is equal to the product of the count rates $n$ of both detectors times the coincidence window. Assuming even photon distribution to both detectors this relation reads: $n_{\mathrm{coinc}} = \Delta \tau \cdot n^2$. The digitisation card is connected to the Computer via a PCI interface. In our setup, correlation histograms are recorded for measurement times of \SI{30}{\s} with a system duty cycle of \SI{31}{\s}. These histograms are individually stored and later added for post-processing and measurement evaluation.

In general terms, a TAC acquisition system offers more degrees of freedom to the user than a standard TDC correlation system, e.g., the bin-width can be tuned. To obtain sufficient counts per bin for correlation experiments with TLS, we chose the TAC readout parameters to yield a bin size $\Delta \tau_{\mathrm{bin}} = \SI{12.2}{\pico \s}$, while employing the highest possible readout noise suppression. To accurately measure the timing resolution of the complete Becker\&Hickl setup, we however changed these parameters to yield a bin size $\Delta \tau_{\mathrm{bin}} = \SI{1.1}{\pico \s}$. Since the TAC detection system lacks channel symmetry and introduces systematics near the TAC range edges, we induce a channel delay between the CFD and the SYNC channel. This channel delay was implemented by a difference in cable length of \SI{3}{m} between both detectors, which together with internal TAC delays and optical setup delays yields a total channel delay of \SI{34.5}{\nano \s}.

\begin{figure}
    \centering
    \includegraphics[width=0.9\columnwidth]{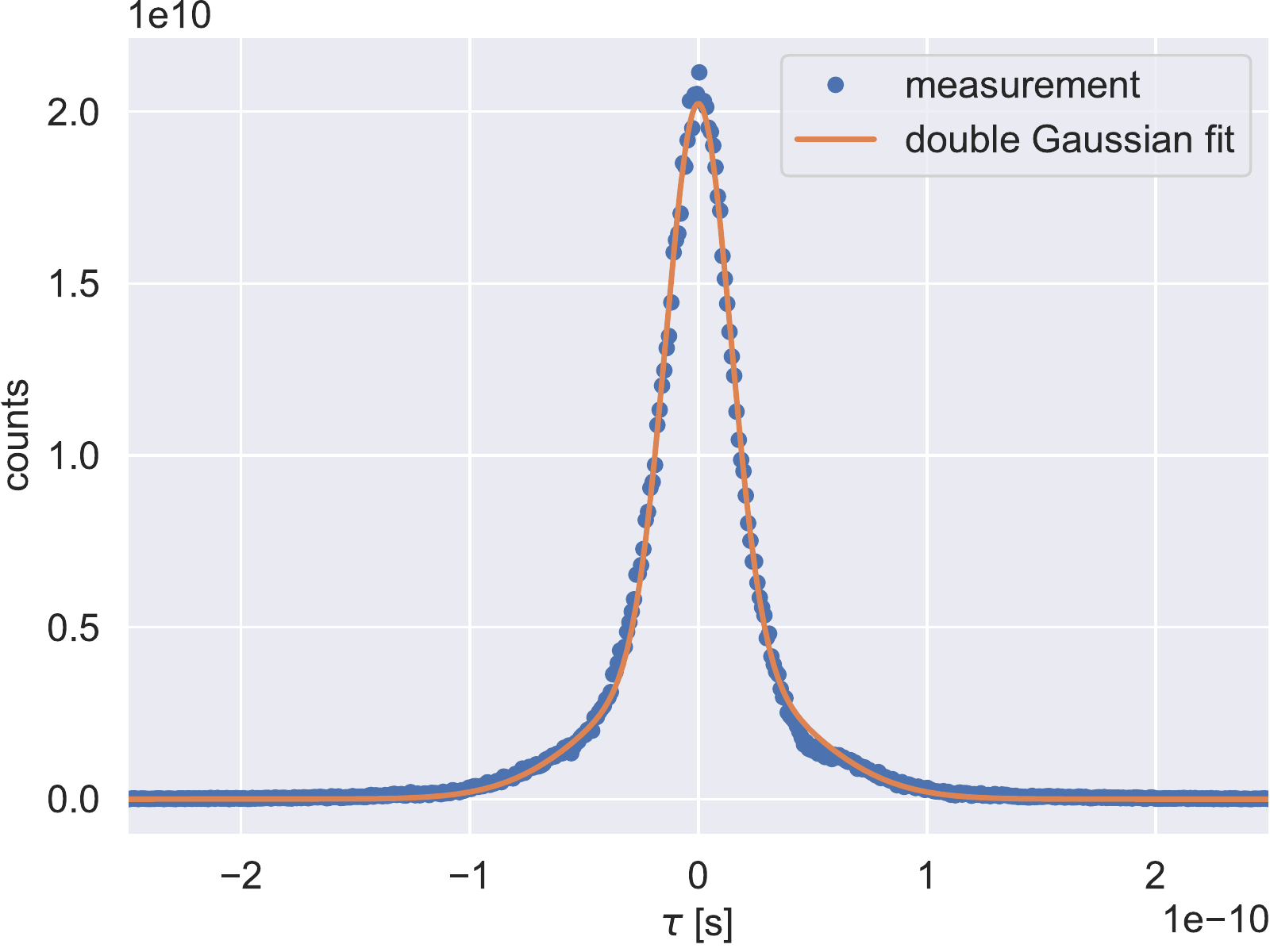}
    \caption{Timing resolution measurement of the complete detection system. The y-axis shows the counts per time bin obtained in the measurement, and the x-axis shows the photon arrival time delay $\tau$ in seconds. We obtain a FWHM timing jitter of \SI[separate-uncertainty = true]{41.61 +- 0.08}{\pico \s}. The measurement data shown as blue dots. Fitting a double Gaussian shown in orange, we can model the timing response of our system to have a wider base and a more narrow peak, which is the main contributor to the short FWHM timing resolution.}
    \label{fig:jitter}
\end{figure}

To determine the timing resolution of the entire detection setup described above without the interference filter, measurements were performed using a femtosecond pulse laser (venteon by Laser Quantum). The resulting correlation histogram is fitted by a double Gaussian function, which fits our measurement well. The full width at half maximum (FWHM) timing jitter was measured to be at maximum $\Delta t_{max} = \SI[separate-uncertainty = true]{41.61 +- 0.08}{\pico \s}$. Our setup thus performs worse than specified by the manufacturer in terms of timing resolution by a factor of 1.4, possibly due to signal broadening in the cables or time delays introduced by the optical elements. An example of a timing resolution measurement and a double Gaussian fit to the measurement results is shown in \cref{fig:jitter}.

\begin{figure}
    \centering
    \includegraphics[width=0.9\columnwidth]{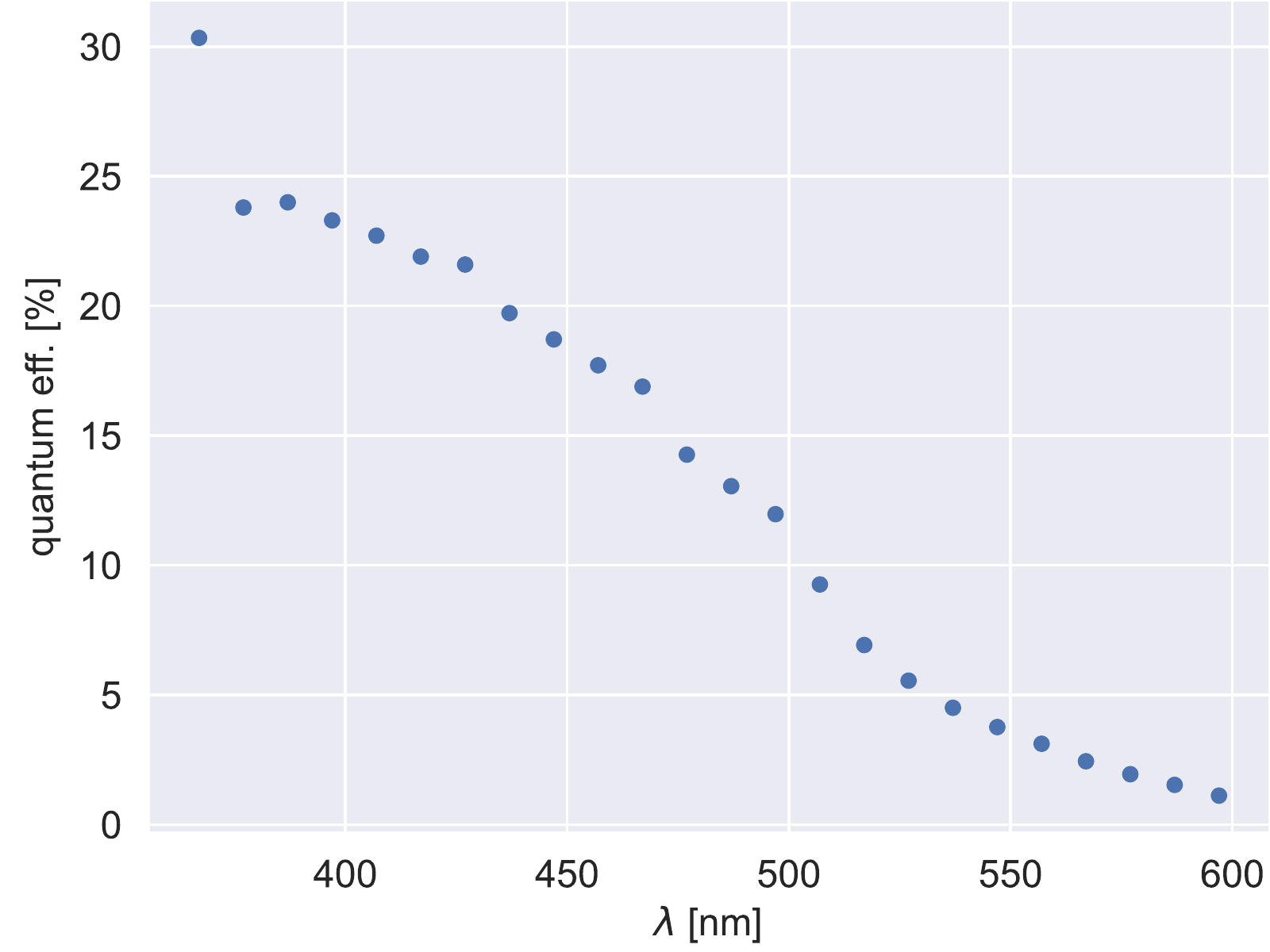}
    \caption{Measured quantum efficiency of a Becker\&Hickl HPM 100-06. The x-axis shows the wavelength in nanometres, the y-axis shows the measured quantum efficiency in per cent. In the blue wavelength region relevant for this paper the quantum efficiency is $\ge 20 \% $.}
    \label{fig:qe}
\end{figure}

The quantum efficiency of the HPDs as a function of wavelength was determined using a high pressure xenon arc lamp (XBO) and a grating based spectrometer. Since timing accuracy is not important in this case, a conventional TDC was used for pulse discretisation and counting. The wavelength dependent quantum efficiency of the Becker\&Hickl HPM 100-06 in the range  \SIrange[separate-uncertainty = true]{380}{600}{\nano \m} is shown in \cref{fig:qe}. It agrees well with the manufacturer specifications, yielding a quantum efficiency of 21.9 per cent at \SI{417}{\nm}.

Concerning systematic uncertainties, the TAC system imprints on the measured correlation histograms in addition to switching systematics at the measurement range edges also more subtle systematics. These include a slope and two superimposed oscillations. Since the amplitude of these systematics is comparable to the measured bunching peak height, these systematics need to be calibrated and compensated for. Such a calibration measurement is performed using scattered light from a halogen light bulb, which due to its ultrashort coherence time produces a bunching peak suppressed by an additional factor $\approx \SI{5e-5}{}$. This calibration measurement is subtracted from the raw measurement to yield a well-resolved correlation peak.

\section{Laboratory test}
\label{sec:lab_test}
We simulated the light of a distant star by utilising a single mode optical fibre to select a single spatial mode emitted by a high-pressure xenon arc lamp (XBO) that resembles an ideal black body radiator. Taking into account the manufacturer-supplied filter transmission spectrum, the shape of the second order correlation function can be calculated according to \cref{eq:g2} and \cref{eq:wct}. This theoretical expectation by construction has a bunching peak height of $g^{(2)}(0) - 1 = 1$, and an extremely short decay-time of the bunching peak on the order of $\tau_c < \SI{1}{\pico \s}$. In order to take into account the timing resolution of our detection system, which is much larger than the coherence time of the light after the filter, this result is convolved with the measured timing jitter (cf. \cref{fig:jitter}). This reduces the height of the bunching peak to its observed value. Finally the measurement expectation needs to take into account the bin size of the measured correlation histogram. While coarser histogram binning enhances the statistics collected for each histogram bin, it also further reduces the bunching peak height from the value obtained from the previously described convolution. For different coincidence rates, slightly different measurement expectations are obtained due to the slightly differing timing jitter. 

\begin{figure}
    \centering
    \includegraphics[width=0.9\columnwidth]{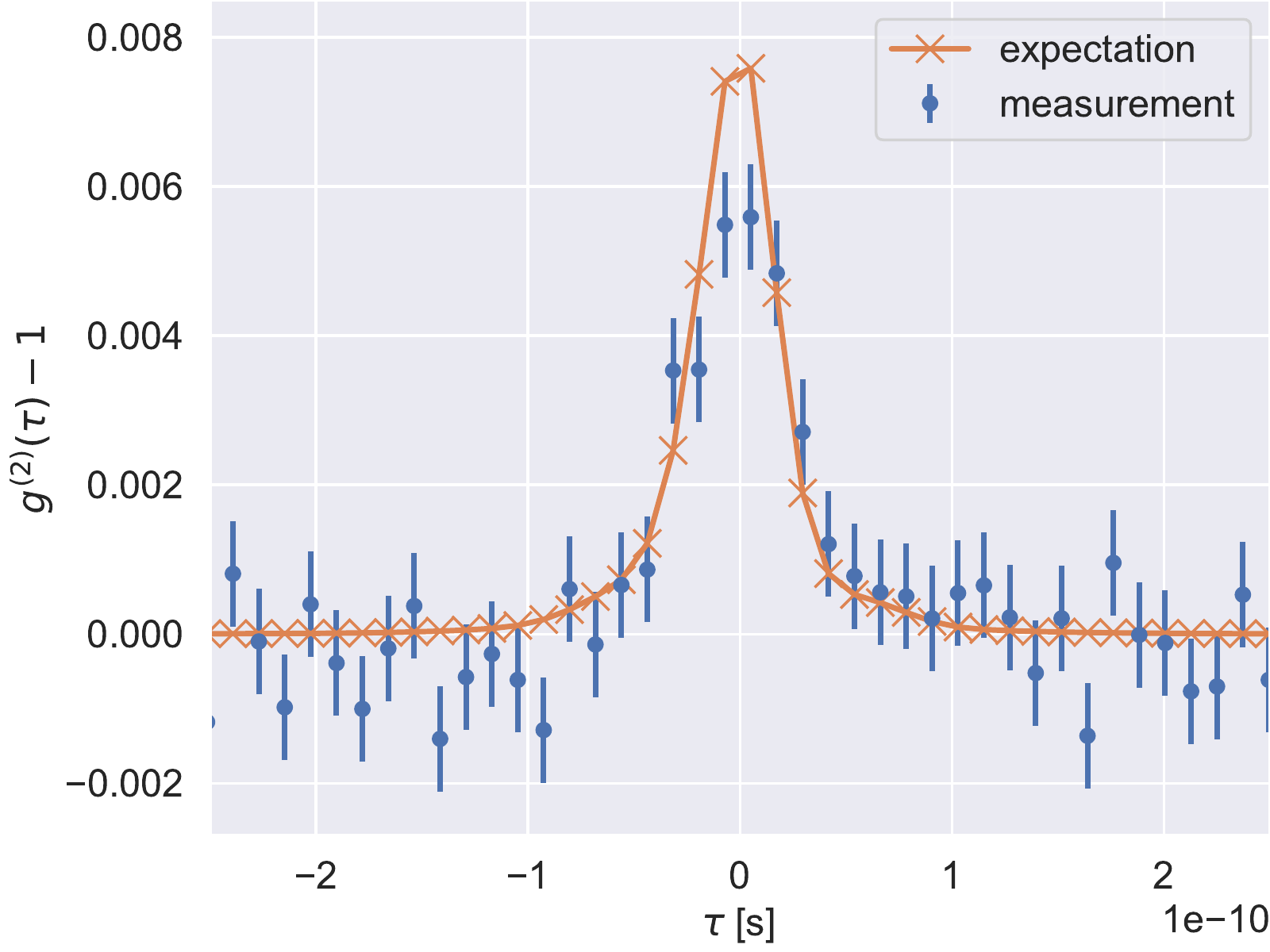}
    \caption{Laboratory test result after an accumulation time of \SI{22.7}{\hour}. The x-axis shows the delay in seconds, while the y-axis shows the second order correlation function with its baseline value of 1 subtracted. The measured coherence time is slightly lower than expected, leading to a smaller bunching peak height in the measurement than in the expectation.}
    \label{fig:lab_test}
\end{figure}

The measurement result shown in \cref{fig:lab_test} is obtained after an accumulation time of \SI{22.7}{\hour} with a mean count rate of approximately \SI{2}{\mega \hertz} per detector. Calculating the correlation peak from the data by numerical integration we recover a coherence time of $\tau_\text{c, meas} = \SI[separate-uncertainty = true]{0.35 +- 0.03}{\pico \s}$, about 85 percent of the coherence time expected from the manufacturer specifications. The expectation for the measurement is additionally displayed in \cref{fig:lab_test}, derived from the expected coherence time of $\tau_\text{c, exp} = \SI{0.425}{\pico \s}$. The error bars shown in \cref{fig:lab_test} correspond to the RMSE taken from the correlation baseline in 20 independent samples of size more than 20 times longer than the FWHM of the measured correlation peak. 

\section{Temporal intensity correlations of Vega}
\label{sec:meas}
Temporal photon correlation measurements of Vega were performed at the \SI{0.5}{\m} Planewave CDK 20 telescope of the Bamberg observatory for six nights in late July and early August of 2020, on days at which weather conditions permitted observation. Between the nights permitting observation, calibration measurements were taken. Measurements at the observatory were generally started at about 10 PM (CEST) and continued until dawn, but measurement beginning and end times vary between measurement days as the optical setup needed to be detached after and reattached before each measurement. An overview of the measurement dates is shown in \cref{tab:campaign}. The table also displays the mean coincidence rate and the percentage of that measurement night to the totally acquired statistics. Until night number 4, the fraction of total statistics gathered per night increased with each night due to the increase in operator experience. Nights number 4, 5, and 6 serve as a realistic benchmark to the performance of our setup, and during these nights higher average coincidence nights could be achieved.

\begin{table}
\centering
\begin{tabular}{c|l|l|l|l}
 night no.  & start & end & $n_{\mathrm{coinc}}$ [Hz] & $p_{\mathrm{tot}}$ \\ \hline
 1 & 2020-07-23T00:33 & 07-23T03:50 & 4832.7 & 8.3\% \\
 2 & 2020-07-29T22:58 & 07-30T04:47 & 3656.8 & 13.1\%\\
 3 & 2020-07-30T23.25& 07-31T04:18  & 4424.5 & 13.3\%\\
 4 & 2020-07-31T22:03& 08-01T04:39  & 5514.3 & 22.4\%\\
 5 & 2020-08-05T21:47& 08-06T04:25 & 5649.9& 23.0\%\\
 6 & 2020-08-06T22:47& 08-07T03:45 & 5764.9 & 20.0\%
\end{tabular}
\caption{Tabular summary of the measurement nights at the Bamberg observatory. In addition to the nights measurement respective start and end timestamps, the average coincidence rate $n_{\mathrm{coinc}}$ obtained during that night as well as the percentage of total coincidence events $p_{\mathrm{tot}}$ gathered during that night is shown. Until night number 4, the fraction of total statistics gathered per night increased with each night due to the increase in operator experience. Nights number 4, 5, and 6 serve as a realistic benchmark to the performance of our setup, and during these nights higher average coincidence nights could be achieved.}
\label{tab:campaign}
\end{table}

\begin{figure}
    \centering
    \includegraphics[width=0.9\columnwidth]{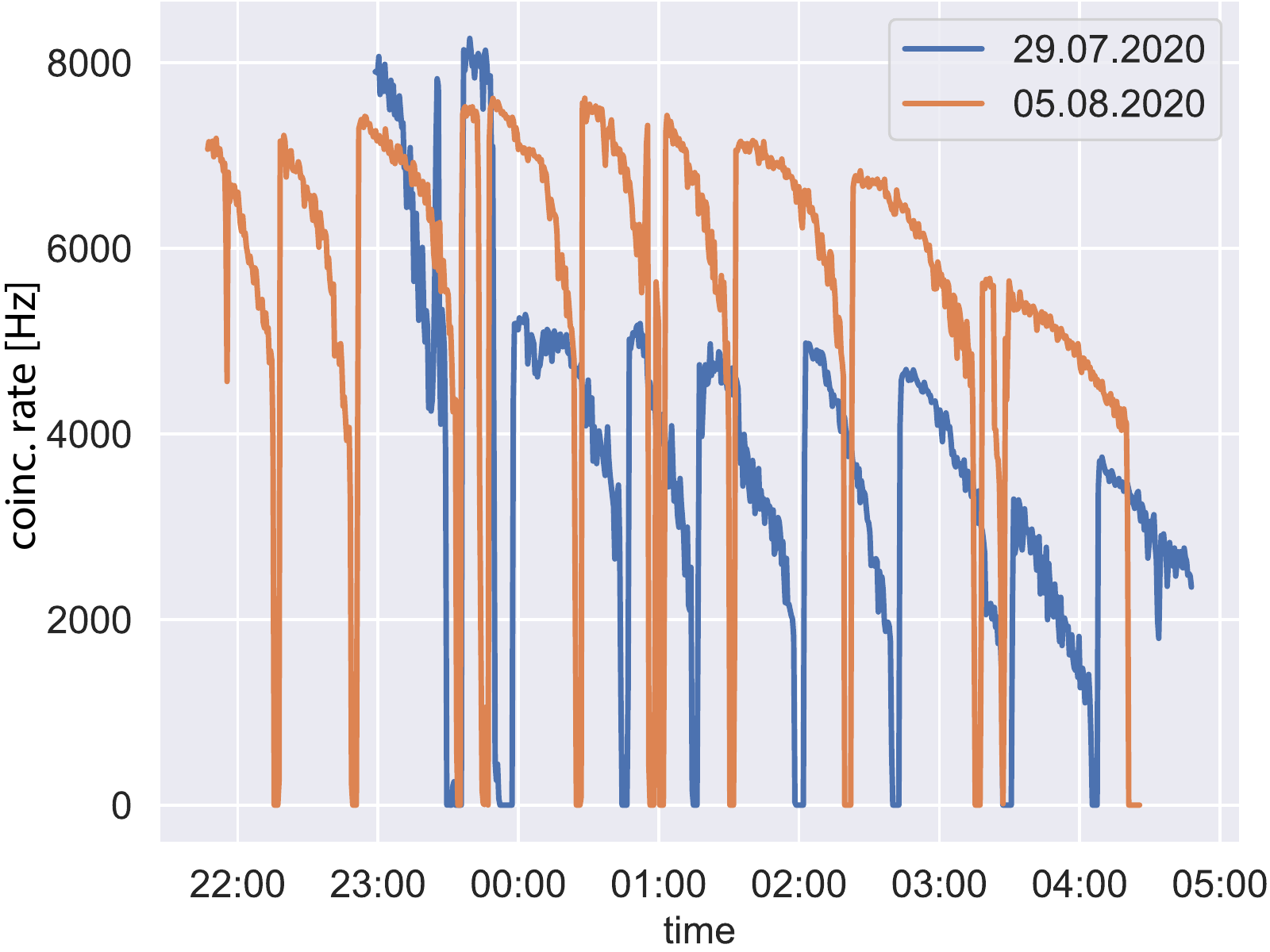}
    \caption{Coincidence rates of two measurement nights. The coincidence rate is depicted on the y-axis, the time is depicted on the x-axis. In the measurement night started 05.08.2020 (shown in orange) the decline of photon detection rate due to the Vega's descent into the atmosphere can be observed despite the zero-rate realignment breaks. The blue curve, related to the measurement started 29.07.2020, shows that the initial coincidence rate could not be fully recovered after the telescope lost the guiding for the second time that night, most likely due to a persisting guiding offset.}
    \label{fig:ADC_rates}
\end{figure}

The photon detection and thus the coincidence rate fluctuated widely during a single measurement night, reaching a maximum of $n_{\mathrm{coinc}} = \SI{8.5}{\kilo \hertz}$ right after optimal alignment and decreasing down to $n_{\mathrm{coinc}}  \approx \SI{0}{\hertz}$ after the guide star had been lost for several minutes, which happened on average every \SI{45}{\min}. \Cref{fig:ADC_rates} shows as an example the coincidence rates of two measurement nights plotted over time. The loss of the guide star and subsequent realignment periods are clearly visible. For realignment, the detectors had to be switched off and the interference filter taken out, but the acquisition electronics was kept on to accurately characterize the entire system performance during the course of the night. In later evaluations, the guiding problem could be traced back to a faulty USB connection between the guiding camera and the telescope control computer. In addition, for every night, a decrease in coincidence rate and thus in photon count rate per detector could be observed as soon as Vega shifted out of its zenith. It should also be noted that in the night starting 2020-07-29, the detected coincidence rate after realignment was significantly lower than before the realignment, and could not be recovered during the course of that night, likely due to a persisting guiding offset.

\begin{figure}
    \centering
    \includegraphics[width=0.9\columnwidth]{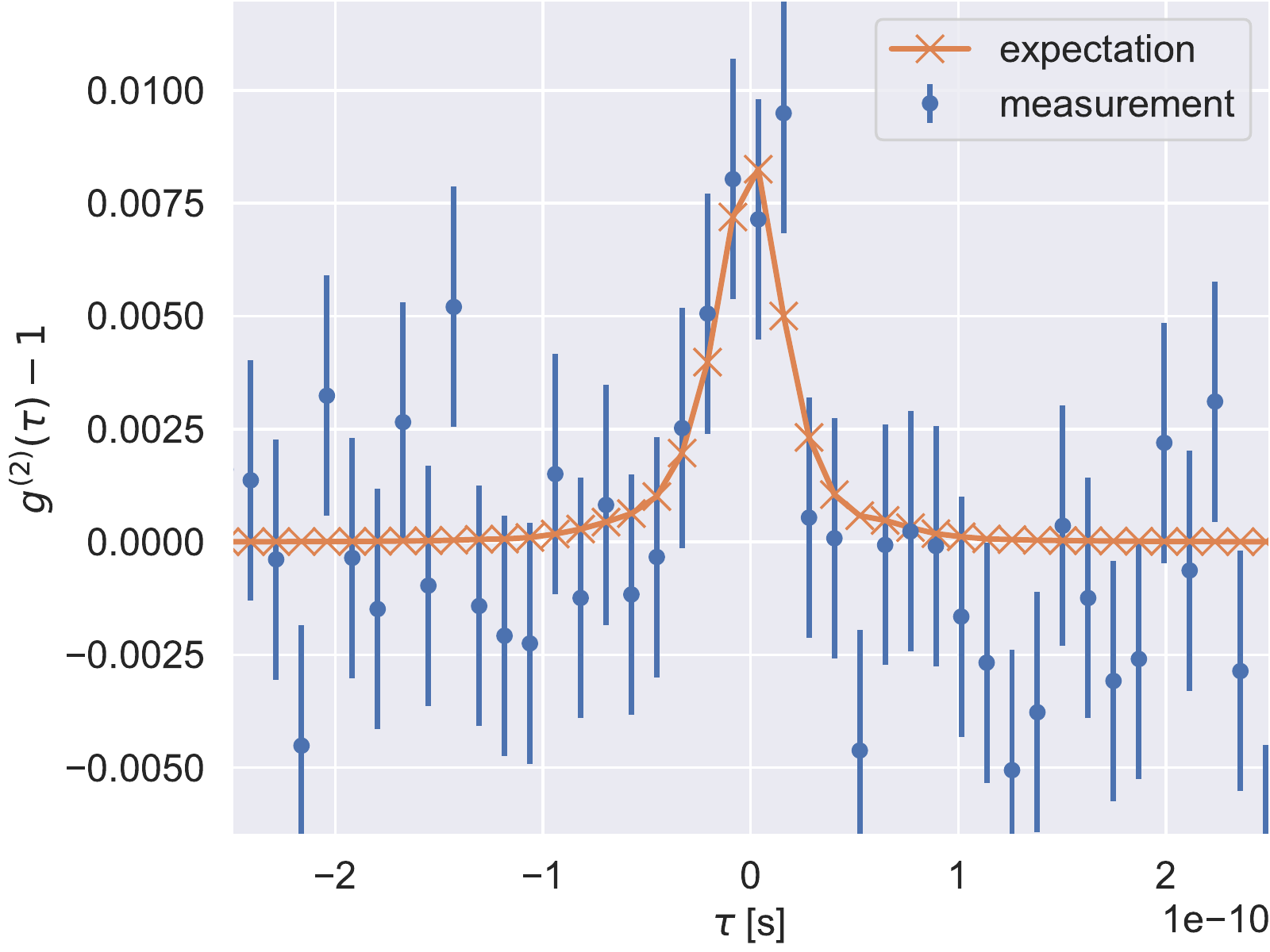}
    \caption{Temporal intensity correlation function of Vega measured with an accumulation time of \SI{32.4}{\hour}. The x-axis shows the temporal delay in seconds, and the y-axis shows the second order correlation function with its baseline value of 1 subtracted. The measured bunching peak shown in blue, is clearly visible around $\tau = 0$, and fits well to the pre-calculated expectation shown in orange.}
    \label{fig:meas_result}
\end{figure}

After a total measurement time of \SI{32.4}{\hour} with a mean coincidence rate of \SI{5.1}{\kilo \hertz} (corresponding to a mean count rate of approximately \SI{337}{\kilo \hertz} per detector) we obtained the measurement result shown in \cref{fig:meas_result}, clearly showing bunching of Vega's starlight in the blue. The error bars are obtained in the same manner as for the laboratory test, using  the calibration measurements recorded between observation nights. Given the final statistics, the measurement result fits the expected result well.

\begin{figure}
    \centering
    \includegraphics[width=0.9\columnwidth]{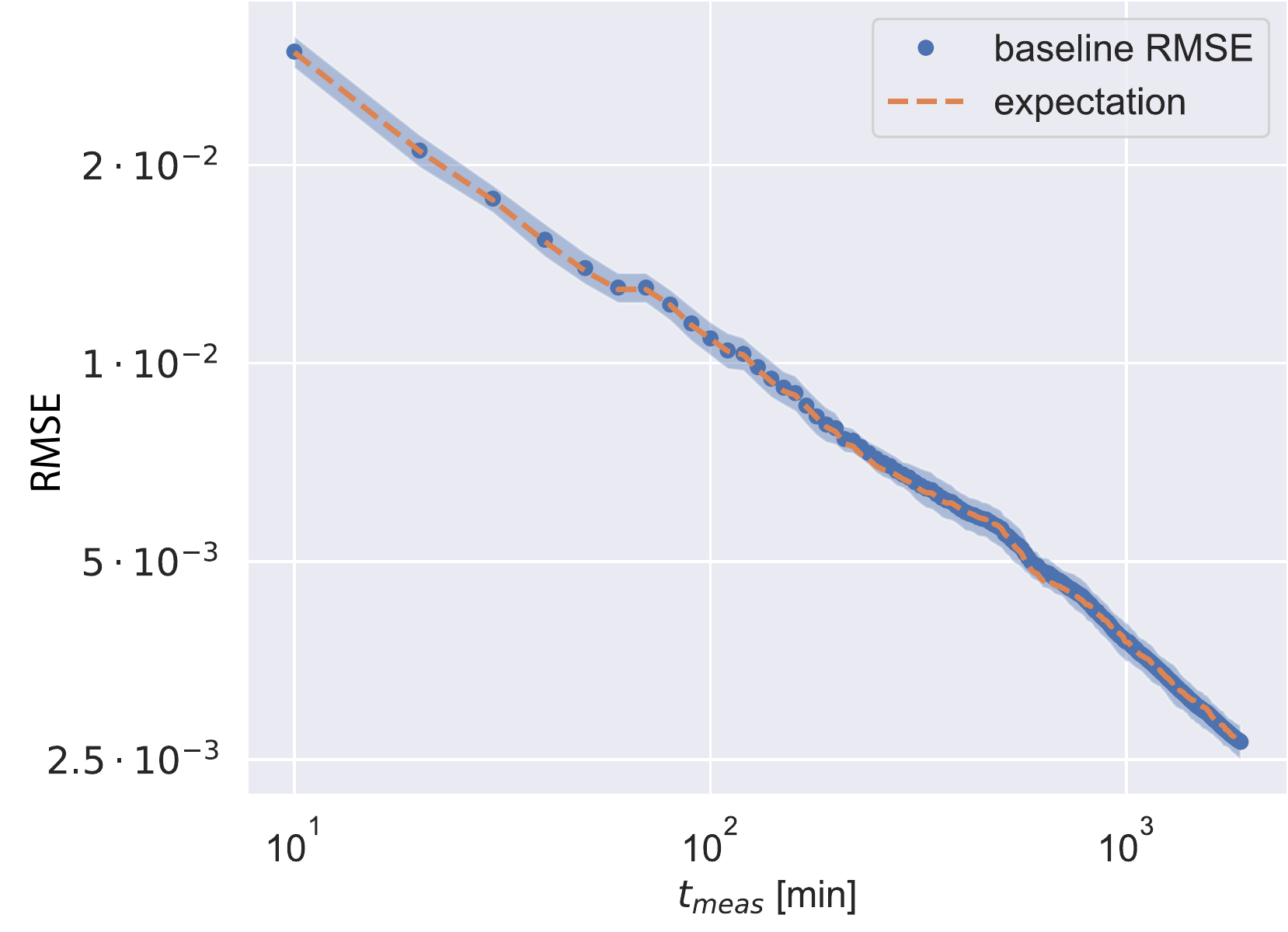}
    \caption{Baseline RMSE evolution over the full measurement time range for the temporal intensity correlation measurement of Vega, whose final result is shown in \cref{fig:meas_result}. The x-axis shows the measurement time in minutes, and the y-axis shows the RMSE calculated from the second order correlation measurements taken up to that time in the manner described in the text. The measured RMSE fits shown as the blue dots has an uncertainty range of the blue shaded region, and fits the shot noise expectation shown in orange very well.}
    \label{fig:RMS_propagation}
\end{figure}

\Cref{fig:RMS_propagation} shows the evolution of the correlation baseline RMSE over the total measurement time together with the shot noise expectation, considering both the statistics of the raw and the calibration histogram. Despite the widely fluctuating count rates our detection system follows the shot noise limit closely, validating our calibration procedure at observatory conditions. Since the RMSE is plotted over the measurement time instead of being plotted over the coincidence counts per bin, it does not form a straight line in the log-log-plot, but rather displays a slope varying with the detector count rates.

Taking into account all efficiencies of the optical components placed within the setup, we expect a total photon detection efficiency of $\sim 9.1 \%$ for each photon arriving at the telescope. Given this efficiency and an average photon flux of Vega of \SI{1400}{\per \second \angstrom \cm^2} in the blue, we calculate the expected photon rate per detector to \SI{638}{\kilo \hertz}. Taking into account the \SI{40}{\nano \s} coincidence window and a maximum \SI{30}{\s} average coincidence rate of \SI{8.5}{\kilo \hertz}, we can estimate the maximum sustained photon rate at our detectors to be \SI{461}{\kilo \hertz}. We note that even this maximal sustained rate is 28 percent lower than the expected rate calculated above, most likely due to additional losses in the telescope optics, especially in  the uncharacterised focuser, and the dense summer atmosphere. Accounting for these additional losses, the overall detection efficiency of our setup is $\sim 6.6 \%$, which is within the range of setups previously reported by \cite{Guerin2017} and \cite{Horch2021}. It should however be noted that this efficiency is only observed right after realignment, as on average only a coincidence rate of \SI{5.1}{\kilo \hertz} corresponding to a photon rate of \SI{382}{\kilo \hertz} per detector was measured. Even accounting for the zero-photon rate brakes for realignment, we suffer from an additional photon detection efficiency decrease of $\sim 14 \%$ for extended measurement times, reducing the overall setup photo detection efficiency to $\sim 5.5 \%$.

After an accumulation time of \SI{32.4}{\hour} the final RMSE was evaluated to \SI{2.7e-3}, with a measured bunching peak height of \SI{9.5e-3} (cf. \cref{fig:meas_result}) , yielding a signal to noise ratio (SNR) of $3.5$. By evaluating the peak integral numerically, a coherence time of $\tau_\text{c, meas} = \SI[separate-uncertainty = true]{0.34 +- 0.12}{\pico \s}$ is obtained, in good agreement with the result derived from the laboratory tests. Note that the bunching peak integral and its error serve as a more reliable metric for estimating the SNR of our bunching measurement, yielding a SNR of 2.8. To our knowledge, this measurement constitutes the first time that single photon counting detectors were used for intensity interferometry in the blue, and the first time that bunching was measured at a telescope of a diameter of merely $\SI{0.5}{\m}$.

\section{Conclusions}
In this paper we have presented results of temporal intensity interferometry with light from Vega using a commercially available Planewave CDK 20 telescope. The required optical components and single photon counting hybrid detectors were all attached directly to the telescope focuser exit. After six nights of measurement, a bunching peak with an amplitude of \SI{9.5e-3}, a coherence time of \SI[separate-uncertainty = true]{0.34 +- 0.12}{\pico \s} and a SNR of $2.8$ was obtained. To our knowledge, this measurement constitutes the first time that single photon counting detectors were used for intensity interferometry in the blue, and the first time that bunching was measured at a telescope of a diameter of merely $\SI{0.5}{\m}$. Our results agree well with both the measurement expectations calculated from the optical and electronic system specifications as well as a laboratory test performed before the astronomical measurement. From the coincidence rates observed at the telescope we calculate the total photo detection efficiency of our full setup to $5.5 \%$. We thus implemented a cost-effective, but sensitive, intensity interferometry experiment using a $\SI{0.5}{\m}$ diameter telescope.

Considering the quantum efficiency of our HPD detectors, we implemented an efficient coupling of the incident starlight to our detection system. In fact, with improved guiding, we expect to be able to transport light from the telescope focus throughout the collimation and filtering stages towards the detectors with hardly any losses. This is mainly due to the large active area of our HPD detectors which makes them insensitive to astronomical seeing. Our setup thus works well at telescopes displaying fast focal ratios. For these telescopes efficient focusing to an area of a few micrometres in diameter, required for fibre coupled superconducting nanowire single-photon detector (SNSPD) or APDs with high timing resolution and thus small active area, is more challenging than for telescopes with larger focal ratio. 

Dobsonian telescopes, which exhibit an even larger focal ratio than the telescope used in this work are commercially available with diameters up to \SI{1}{\m}. Expanding our setup to two \SI{1}{\m} Dobsonian telescopes, would increase our sensitivity by a factor of $(2\times2)^2 = 16$, enabling observations of spatial correlations of mag. 0 stars with $5 \sigma$ in about \SI{3}{\hour}. For this estimation, we use the fact that the SNR acquired per time interval is proportional to the count rate squared \citep{Brown1957}, and the detected rate is proportional to the area of the telescope, whose diameter we doubled. If, as suggested in \cite{Trippe2014}, highly flexible arrays of movable Dobsonian telescopes are used in combination with colour multiplexing with a modest number of central wavelengths within one band, the accumulation time for the observation of spatial correlations of a magnitude 2 star with $5 \sigma$ could be lowered to less than \SI{3}{\hour}. Such an array of large, wavelength multiplexed Dobsonians would serve as a cost-effective way of implementing an intensity interferometer. First steps into this direction have already been taken by \cite{Horch2021}. 

Mirrors even larger in diameter are currently already in use for intensity interferometry experiments at IACTs. Due to the high fluxes obtained at these large telescopes, these experiments currently rely on correlating the photo-current of photomultiplier tubes. Recently, prototypes single photon counting detectors with up to \si{\giga \hertz} burst rates, large areas, low timing jitter, and high quantum efficiencies have been realised by \cite{Orlov2019a}. The use of such detectors could be especially beneficial at telescopes like MAGIC or the Cherenkov Telescope Array's Large Size Telescopes, which have an intrinsic telescope timing jitter on the order of \SI{100}{\pico \s} \citep{Shayduk2005}. 

\section*{Acknowledgements}

We thank Verena Leopold and a previous reviewer for their comments and careful proof-reading of this paper. Further, we thank Vahid Sandoghdar for his support in acquiring the detection system, and Oleg Kalekin as well as Tobias Boolakee for support in system calibrations. For assistance during the measurement nights at the Bamberg observatory we are grateful to the entire Bamberg observatory staff,  especially to  Uli Heber, J\"oern Wilms, Philipp Weber, and Eva von Gem\"unden. Finally we want to thank Andreas Zmija, Naomi Vogel, Adrian Zink,  Gisela Anton, and Stefan Funk for their collaboration in the field of intensity interferometry and for helpful discussions on this topic.

\section*{Data Availability}
The data directly supporting the plots is available at \url{https://doi.org/10.22000/1559}. Intermediary correlation histograms are available in time-intervals of \SI{30}{\s} upon reasonable request from the corresponding author. Due to the nature of the correlation process no raw photon event stream can be supplied.



\bibliographystyle{rasti}
\bibliography{full_bib} 





\bsp	
\label{lastpage}
\end{document}